\documentclass[aps,prl,twocolumn,showpacs,amsmath,amssymb,superscriptaddress]{revtex4} 
\usepackage{graphicx} 
\usepackage{amssymb} 
\usepackage{natbib} 
\usepackage{color}
\usepackage{epstopdf}

\begin{document}

\title{Stabilization of $s$-wave superconductivity through arsenic $p$-orbital
	hybridization in electron-doped BaFe$_2$As$_2$}

\author{David W. Tam} \affiliation{Department of Physics and Astronomy, Rice
	University, Houston, Texas 77005, USA}

\author{Tom Berlijn} \affiliation{Center for Nanophase Materials Sciences and
	Computational Sciences and Engineering Division, Oak Ridge National
	Laboratory, Oak Ridge, Tennessee 37831-6494, USA}

\author{Thomas A. Maier} \email{maierta@ornl.gov} \affiliation{Center for
	Nanophase Materials Sciences and Computational Sciences and Engineering
	Division, Oak Ridge National Laboratory, Oak Ridge, Tennessee 37831-6494,
	USA}

\date{\today} 
\begin{abstract}
	
	Using random-phase approximation spin-fluctuation theory, we study the
	influence of the hybridization between iron $d$-orbitals and pnictide
	$p$-orbitals on the superconducting pairing state in iron-based
	superconductors. The calculations are performed for a 16-orbital
	Hubbard-Hund tight-binding model of BaFe$_2$As$_2$ that includes the
	As-$p$ orbital degrees of freedom in addition to the Fe-$d$ orbitals and
	compared to calculations for a 10-orbital Fe-$d$ only model.  In both
	models we find a leading $s^\pm$ pairing state and a subleading $d_
	{x^2-y^2}$-wave state in the parent compound. Upon doping, we find that
	the $s^\pm$ state remains the leading state in the 16-orbital model up to
	a doping level of 0.475 electrons per unit cell, at which the hole Fermi
	surface pockets at the zone center start to disappear. This is in contrast
	to the 10-orbital model, where the $d$-wave state becomes the leading
	state at a doping of less than 0.2 electrons. This improved stability of
	$s^\pm$ pairing is found to arise from a decrease of $d_{xy}$ orbital
	weight on the electron pockets due to hybridization with the As-$p$
	orbitals and the resulting reduction of near $(\pi,\pi)$ spin-fluctuation
	scattering which favors the competing $d$-wave state. These results show
	that the orbital dependent hybridization of Fermi surface Bloch states
	with the usually neglected $p$-orbital states is an important ingredient
	in an improved itinerant pairing theory.
\end{abstract}

\maketitle

\section{Introduction}

The detailed nature of the pairing mechanism that gives rise to
superconductivity in the iron-based superconductors continues to be a matter
of debate. Based on the proximity of the superconducting state to the magnetic
stripe order observed in most parent and weakly doped materials,
antiferromagnetic spin fluctuations have been widely discussed to play a major
role. Early on it was predicted that these fluctuations, occurring at a
wavevector $Q=(\pi,0)$  that separates the Fermi surface hole pockets at the
zone center and the electron pocket at $(\pi,0)$, will mediate an $s^\pm$
superconducting state, in which the gap changes sign between the hole and
electron pockets \cite{MSJD08}.

Given the metallic character of the iron-based parent compounds, a
weak-coupling fluctuation exchange picture is a natural platform for
understanding superconductivity in these systems. Random phase approximation
(RPA) based spin-fluctuation calculations for realistic tight-binding models
of these systems indeed find an $s^\pm$ superconducting state \cite{Hirschfeld15}. However,
these calculations also suggest a strongly competitive $d_{x^2-y^2}$ pairing
channel \cite{Kuroki08,GMHS09}, which can even become dominant with either electron or hole
doping. In fact, for certain cases, this transition from an $s$-wave to a
$d$-wave ground state happens already at very small levels of electron
doping. Fig.~1, for example, displays the results
of an RPA calculation for a three dimensional (3D) 10-orbital model (2 Fe per
unit cell) of BaFe$_2$As$_2$, where this change already occurs at a doping
level of 0.1 electrons per Fe. Experimentally, the existence of an $s$-wave
superconducting gap in doped BaFe$_2$As$_2$ is broadly supported by
experiments including ARPES, muon spin relaxation, optical reflectivity, heat
capacity, neutron scattering, and other techniques
\cite{HiKM11,TSBN09,John10,ABBC08}. However, there is no evidence for a change
to $d$-wave gap symmetry at small doping levels.

This problem of spin-fluctuation RPA theory is intimately linked to the momentum
structure of the spin-fluctuation interaction. For systems with both hole- and
electron pockets of approximately equal size such as in the parent compounds,
the RPA spin susceptibility is dominated by a strong peak at $Q=(\pi,0)$, which
arises from the nesting between the hole pockets and the electron pockets and
which favors the $s^\pm$ state. But additional scattering between opposite sides
of the electron pockets gives rise to a ridge-like structure around $(\pi,\pi)$,
which becomes dominant already at small levels of electron doping and which
favors the $d$-wave state \cite{Fernandes13}.

The low-temperature phase of optimally superconducting BaFe$_2$As$_2$
crystallizes with tetragonal $I4/mmm$ symmetry containing an inversion center.
This gives rise to two distinct iron sublattices and, as a result, a model
containing ten iron-$d$ orbitals (five each for the two inequivalent iron
sites in the unit cell) is the minimal model required to generate the
appropriate orbital eigenstates at the Fermi energy.

In this work, we show that by including the six inequivalent arsenic $p$
orbitals in addition to the ten iron $d$ orbitals, the competing $d$-wave
pairing state is suppressed and the $s^\pm$ state remains the leading pairing
state up to an electron doping level where the hole pocket disappears and low
energy spin-fluctuations near $(\pi,0)$ are suppressed. These results suggest
that the hybridization of the $d$-orbitals with the arsenic $p$ states,
despite their relatively small spectral weight at the Fermi energy, is an
important factor in a complete picture of superconductivity in these
materials.

\section{Method}

Iron in BaFe$_2$As$_2$ is nominally in the 2+ oxidation state, leaving a
$3d^6$ shell (12.0 carrier electrons per two-iron unit cell); optimal SC
occurs at BaFe$_{1.9}$Ni$_{0.1}$As$_2$, corresponding to a donation of two
electrons per $3d^8$ Ni (12.2 carriers), while SC is extinguished near
BaFe$_{1.75}$Ni$_{0.25}$As$_2$ (12.5 carriers). For the 16-orbital models, As
takes the 3- state, yielding $4p^6$ and contributing an extra 12 carriers per
unit cell; BaFe$_{1.9}$Ni$_{0.1}$As$_2$ then corresponds to 24.2 carriers.
We use the WIEN2K software package \cite{wien2k} to generate multi-orbital
tight-binding models for the parent (undoped) compounds using the experimentally-determined lattice positions of
the As ions, which we project into a tight-binding Wannier basis with
WIEN2WANNIER \cite{wien2wannier} and WANNIER90 \cite{wannier90}.
From these models, our methods proceed as described before
\cite{Kuroki08,GMHS09,MaSc08,MGSH09,MGHS11,MaHS12,WKZB13}, first calculating the
bare magnetic susceptibility tensor

\begin{widetext}
	\begin{equation}\label{eqn:chi0} 
		\begin{aligned}
			\chi^0_{\ell_1\ell_2\ell_3\ell_4} (\mathbf{q},\omega) =
			- \frac{1}{N} \sum_{\mathbf{k},\mu\nu} \
			\frac{a^{\ell_4}_\mu(\mathbf{k})
			a^{\ell_2,*}_\mu(\mathbf{k})
			a^{\ell_1}_\nu(\mathbf{k}+\mathbf{q})
			a^{\ell_3,*}_\nu(\mathbf{k}+\mathbf{q})}
			{\omega+E_\mu(\mathbf{k}) - E_\nu(\mathbf{k}+\mathbf{q})
			+ i\delta} \ \left[ f(E_\mu(\mathbf{k}),T) -
			f(E_\nu(\mathbf{k}+\mathbf{q}),T) \right] \qquad , 
		\end{aligned}
	\end{equation}
\end{widetext}
with the band indices $\mu$ and $\nu$, and orbital indices $\ell_i$. The
matrix elements $a^\ell_\mu(\mathbf{k}) = \langle \ell|\mu\mathbf{k} \rangle$ represent
the orbital projection of the Bloch states, $f(E_\mu({\bf k}),T)$ is the Fermi function for band energy $E_\mu({\bf k})$ at
temperature $T$ which we set to $100$ K. For the sum over $\mathbf{k}$ we use a mesh of
40$\times$40$\times$4 points over the 3D Brillouin zone. We then compute the
RPA spin and charge susceptibility tensors
\begin{align}
	\chi^{s/c,{\rm RPA}}_{\ell_1\ell_2\ell_3\ell_4}({\bf q},\omega) &=
	\left\{ \chi^0({\bf q},\omega)[1-\mathcal{U}^{s/c}\chi^0({\bf
	q},\omega)]^{-1} \right\}_{\ell_1\ell_2\ell_3\ell_4} 
\end{align}
using the interaction matrices in orbital space for the spin ($\mathcal{U}^s$)
and charge ($\mathcal{U}^c$) channels, which contain linear combinations of
intra- and inter-orbital Coulomb repulsions $U$ and $U'$, respectively, as
well as Hund's rule and pair-hopping terms $J$ and $J'$, respectively \cite{GMHS09}. We
have used spin rotational invariant combinations that satisfy $U'=U/2$ and
$J=J'=U/4$ \cite{NoRo14}.
We find that adjusting the ratios of the Coulomb interaction parameters, such as the choice $J=U/10$, has no effect
on the conclusions of this work, except to scale the RPA susceptibility and pairing eigenvalues.
We choose different values for $U$ for the 10- and
16-orbital models (keeping the parameter ratios fixed) as explained below. The
physical spin susceptibility is then given by
\begin{align}
	\chi^{s,{\rm RPA}}({\bf q},\omega) =
	\frac{1}{2}\sum_{\ell_1\ell_2}\chi^{s,{\rm
	RPA}}_{\ell_1\ell_1\ell_2\ell_2}({\bf q},\omega)\,. \label{eq:SpinSus} 
\end{align}

The superconducting properties are calculated from the pairing vertex in
band representation 
\begin{align}
	\label{eqn:gammapp} \Gamma_{ij}(\mathbf{k},\mathbf{k'}) = \text{Re}
	&\sum_{\ell_1 \ell_2 \ell_3 \ell_4} \ a^{\ell_1,*}_{\nu_i}(\mathbf{k})
	a^{\ell_4,*}_{\nu_i}(-\mathbf{k}) \\ &\times \Gamma_{\ell_1 \ell_2
	\ell_3 \ell_4}(\mathbf{k},\mathbf{k'},\omega=0)
	a^{\ell_2}_{\nu_j}(\mathbf{k'}) a^{\ell_3}_{\nu_j}(-\mathbf{k'})\,,
	\nonumber 
\end{align}
where the momenta ${\bf k}\in {\cal C}_i$ and ${\bf k'}\in {\cal C}_j$ are
restricted to the electron and hole Fermi surface sheets ${\cal C}_{i/j}$ and
$\nu_{i/j}$ are the band indices of these sheets. The scattering vertex
$\Gamma_{\ell_1\ell_2\ell_3\ell_4}$ describes the particle-particle scattering
in orbital space and is given in RPA approximation as
\begin{align}
	\label{eqn:singletvertex} \Gamma_{\ell_1 \ell_2 \ell_3
	\ell_4}(\mathbf{k},\mathbf{k'},\omega) &= \left[\frac{3}{2}
	\mathcal{U}^s \chi^s_\text{RPA}(\mathbf{k}-\mathbf{k'},\omega)
	\mathcal{U}^s \right.\\ + \frac{1}{2} \mathcal{U}^s &-\left.\frac{1}{2}
\mathcal{U}^c \chi^c_\text{RPA}(\mathbf{k-k'},\omega) \mathcal{U}^c +
\frac{1}{2} \mathcal{U}^c \right]_{\ell_1\ell_2\ell_3\ell_4}\,.\nonumber 
\end{align}
The momentum structure $g(k)$ of the pairing state can then be found by solving
the eigenvalue problem \cite{GMHS09} 
\begin{align}
	- \sum_{j} \oint_{C_j} \frac{d{\bf k}'_{\parallel}}{2\pi v_F({\bf
	k}'_\parallel)} \Gamma_{ij}(\mathbf{k},\mathbf{k'})
	g_\alpha(\mathbf{k'}) = \lambda_\alpha g_\alpha(\mathbf{k})
	\label{eqn:eigenvalueeqn} 
\end{align}
where the eigenfunction $g_\alpha({\bf k})$ corresponding to the largest
eigenvalue $\lambda_\alpha$ gives the leading pairing instability of the
system.
We calculate the pairing states for different electron dopings by applying a rigid band shift.

\section{Results}

For the undoped case (12 electrons per unit cell), we generally find a leading
s-wave solution with $s^\pm$ structure where the gap changes sign between hole
and electron pockets. In both the 10- and 16-orbital models, the gap is fairly
isotropic on the hole pockets and displays similar angular variation on the
electron pockets but is nodeless. The second leading solution we observe in
the 10-orbital model has $d_{x^2-y^2}$ structure with nodes on both the hole
and electron pockets (due to hybridization of the latter). In the 16-orbital
model, we also observe this $d_{x^2-y^2}$ state, but, in addition, there are
other $d$-wave solutions (including $d_{xy}$) with larger eigenvalues. Since
these other states rapidly disappear with electron doping we ignore them in the
following discussion and focus on the doping dependence of $s^\pm$ and $d_
{x^2-y^2}$-wave states.

\begin{figure}[ht!]
	\includegraphics[scale=.4]{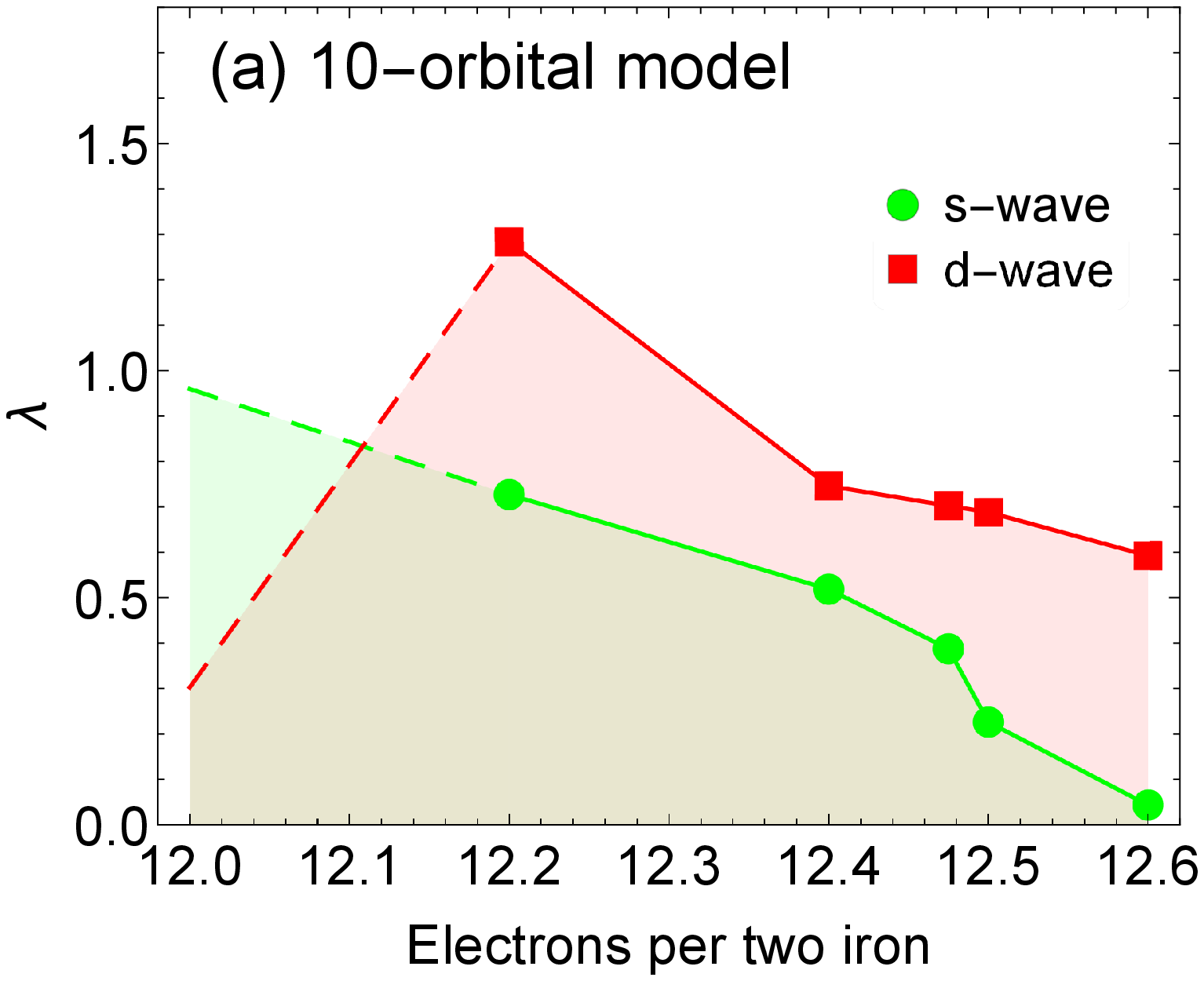} 
	\includegraphics[scale=.4]{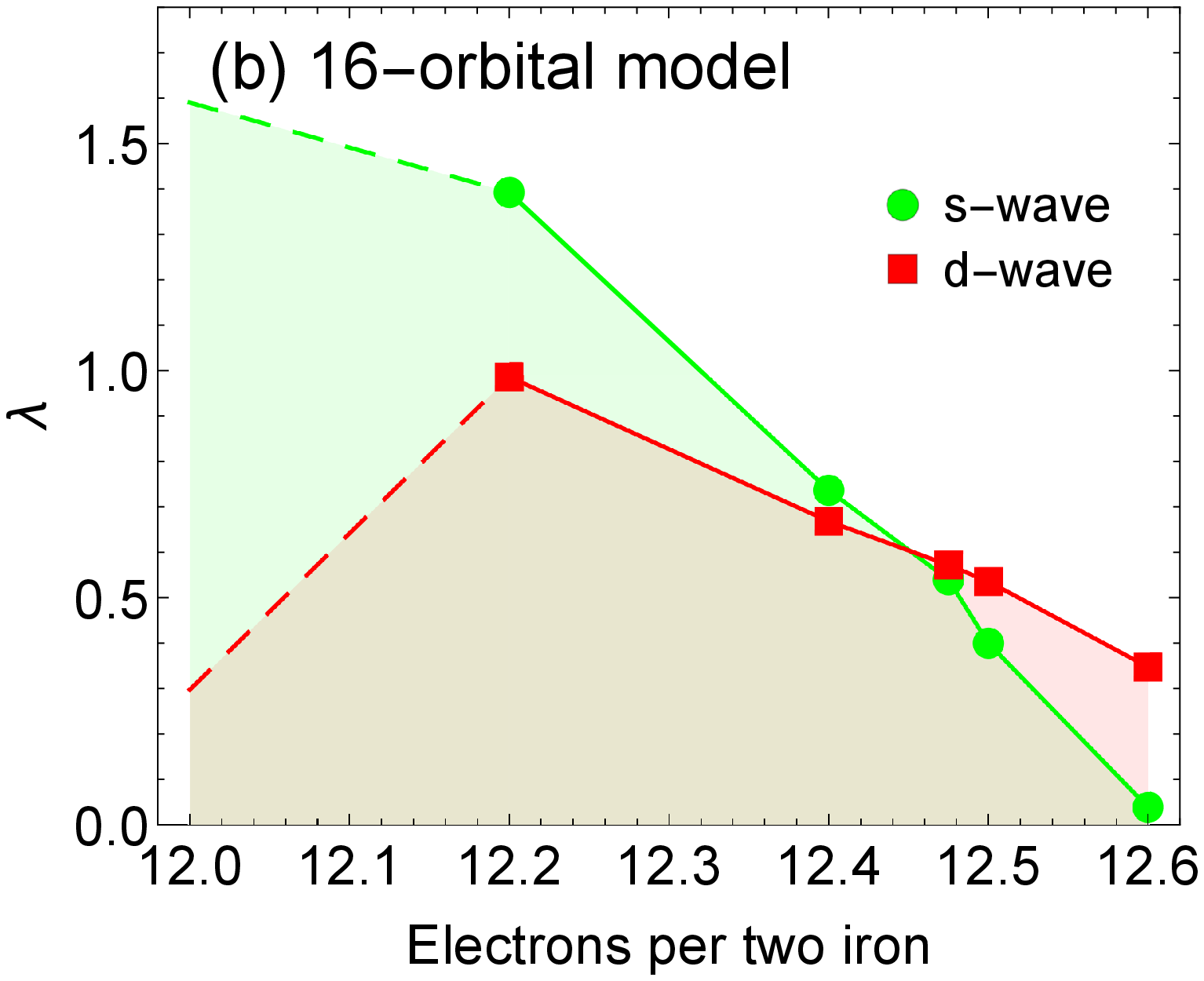} 
	\caption{ Pairing eigenvalues for the 10-orbital (a) and 16-orbital model
		(b) for electron-doped BaFe$_2$As$_2$ using a rigid band shift. The parent compound has
		$\langle n \rangle$ = 12.0 electrons per two iron, whereas $\langle n
		\rangle$ =12.5 corresponds to a nominal doping of
		BaFe$_{1.75}$Ni$_{0.25}$As$_2$ or BaFe$_{1.5}$Co$_{0.5}$As$_2$. The
		dashed lines show that different RPA interactions are used for the
		undoped compounds that give eigenvalues close to 1. } \label{fig1}
\end{figure}

Fig.~1 shows the doping dependence of the $s^\pm$ and $d$-wave eigenvalues for
the 10-orbital model, which does not include the arsenic orbitals, and the
full 16-orbital model that includes the three (per arsenic) $p$-orbitals. In
the moderate to strong doping regime (12.2/24.2 electrons per unit cell and
higher), the RPA interactions are chosen to generate eigenvalues roughly in
the range 0 to 1; we use $U=1.10$ and $1.72$ for the 10- and 16-orbital
models, respectively. The fact that a larger interaction $U$ needs to be
chosen to give similar pairing strengths in the 16-orbital model reflects the
fact that the additional screening from the $p$ orbitals is not included in
the bare interaction parameters of the 16-orbital models. Using the same
interaction parameter $U$ for the parent compound (12.0/24.0 electrons per
unit cell) would lead to an antiferromagnetic spin-density wave instability at
$ (\pi,0)$. In order to suppress this instability and study the order of the
leading pairing solutions in this regime, we choose a different $U$ for the
undoped case. For the 10-orbital model we set $U=0.92$ and choose 1.40 for the
16-orbital model. This choice gives the same $d$-wave eigenvalues
$\lambda_d=0.3$ for both models.

\begin{figure*}[ht]
	\includegraphics[scale=.6]{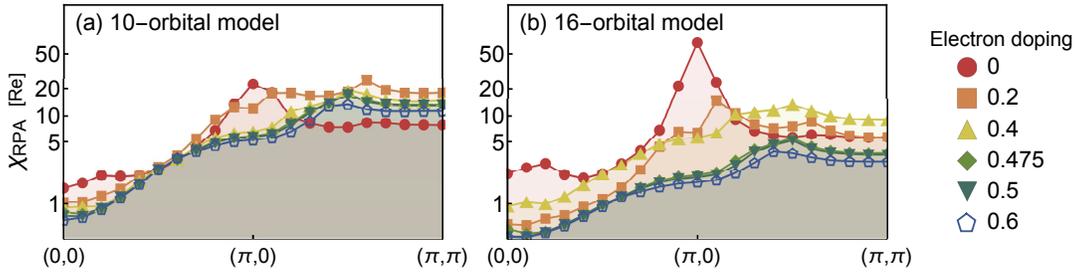} 
	\caption{ (Note log scale.) Cuts of the RPA
		susceptibility along high-symmetry directions in the 1 Fe/unit
		cell Brillouin zone, for the electron-doped system, and for the
		10-orbital (a) and 16-orbital (b) models. The RPA interaction
		$U$ is chosen just below the critical strength that leads to a
		divergence of $\chi_\text{RPA}$. This divergence occurs at U=1.19 for the 10-orbital model, and 1.73 for the 16-orbital model, for 12.2 electrons per iron in both cases. For the parent compounds, interactions are chosen to give pairing eigenvalues close to 1.
	} \label{fig2} 
\end{figure*}

With this choice of interaction parameters, we find that the $s^\pm$ state is
the leading instability in the undoped case for both models, but the margin by
which it is leading over the $d$-wave state is much larger in the 16-orbital
model than in the 10-orbital model. Most importantly in the 16-orbital model the
$s^\pm$ state remains the leading state up to a doping of approximately 0.475
electrons per unit cell. In contrast, in the 10-orbital model, the $d$-wave
state becomes the leading state already at a doping of less than 0.2 electrons
per unit cell.

In order to understand this significant increase in the stability of the
$s^\pm$ state over the $d$-wave state in the 16-orbital model, we now examine
the momentum structure of the zero frequency RPA spin susceptibility
$\chi^{s,\text{RPA}} ({\bf q},\omega=0)$ (Eq.~\eqref {eq:SpinSus}), which
enters the paring interaction Eq.~\eqref{eqn:singletvertex}. Fig.~\ref {fig2}
shows $\chi^{s,\text{RPA}} ({\bf q},\omega=0)$ for ${\bf q}$ along high
symmetry directions in the 1 Fe/unit cell Brillouin zone for different dopings
in the 10-orbital (a) and 16-orbital (b) models. As noted before, the
spin-fluctuation scattering at $(\pi,0)$ favors the $s^\pm$ pairing state,
while the $d$-wave state arises from scattering near $(\pi,\pi)$. For the
undoped case, one sees a peak at $(\pi,0)$ for both the 10-orbital and the
16-orbital models, leading to the dominant $s^\pm$ state that is found in
Fig.~\ref{fig1}. However, this peak is much more enhanced over the scattering
near $(\pi,\pi)$ in the 16-orbital model than in the 10-orbital case. This
explains the larger difference between the $s^\pm$ and the $d$-wave
eigenvalues in the 16-orbital model at half-filling. As the doping increases,
one sees that the $(\pi,0)$ peak relative to the $(\pi,\pi)$ peak diminishes slower in the 16-orbital model than in the 10-orbital model. For example at $x=0.2$  the susceptibility maximum is no longer at $(\pi,0)$ in the 10-orbital model, whereas in the 16-orbital model it still is. As a result, the $s^\pm$ state remains more
stable with increasing doping and dominant over the $d$-wave up to a doping
level of $\sim 0.475$ electrons.

\begin{figure} [t] 
\includegraphics[scale=.45]{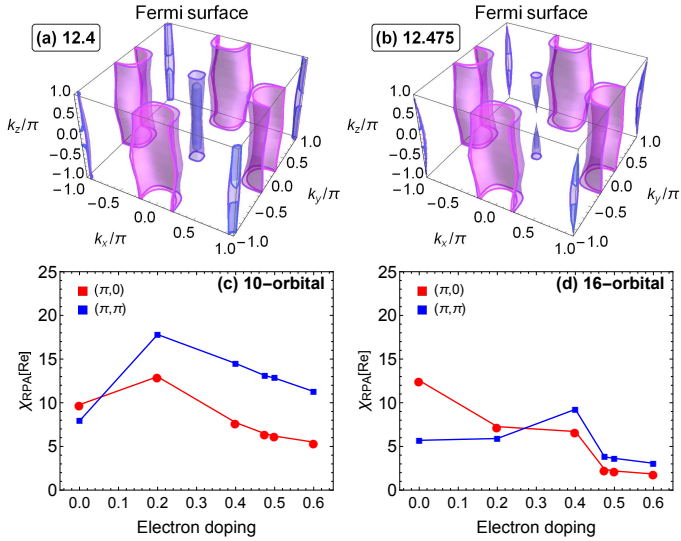} 
	\caption{Fermi surface in the 16-orbital model for a filling of 12.4 electrons 
	(a) and 12.475 electrons (b) per unit cell. Doping dependence of the RPA
	spin susceptibility $\chi^{s,\text{RPA}}({\bf q},\omega=0)$ for ${\bf q}$
	near $(\pi,0)$ and $(\pi,\pi)$ in the (c) 10-orbital and (d) 16-orbital
	model. The Fermi surfaces in (a) and (b) are for the 16-orbital model but are found
	to be indistinguishable from those in the 10-orbital model.}
	\label{fig3} 
\end{figure}

This doping corresponds exactly with the filling where the Fermi surface hole
pockets at the zone center start to disappear.  Fig.~3  shows the Fermi
surface of the 16-orbital model for a filling of 12.4 (a) and 12.475 (b) and
one sees that the hole pockets at the zone center are disappearing for a
filling of 12.475. The fact that the $d$-wave pairing state becomes leading at
this doping is a consequence of our Fermi surface restricted treatment of the
pairing problem, since the low energy spin fluctuations near $(\pi,0)$ that
favor the $s^\pm$ state are suppressed when the hole bands move below the
Fermi energy.  This is shown in the bottom panel in Fig.~3, where we plot the
real part of the RPA spin susceptibility $\text{Re}\,\chi^{s,\text{RPA}}({\bf
q},\omega=0)$ for different dopings. Here we have integrated the
susceptibility over small 2D regions of size $(0.2\pi,0.2\pi)$ around the nominal position,
and averaged over $q_z$, although we find little difference for any single
$q_z$. In the 10-orbital model, upon doping, the $(\pi,\pi)$ scattering
immediately becomes much stronger than the $(\pi,0)$ scattering. In contrast,
in the 16-orbital model the $(\pi,0)$ scattering remains dominant up to a doping of 0.25
electrons, and for higher doping the two regions show very similar magnitude.
We also find that the spin gap at $(\pi,0)$ in the imaginary part of $\chi({\bf q},\omega)$ also increases
significantly upon crossing the threshold of 12.475 electrons, and spectral
weight is transferred to higher energies upon further doping. This is similar
to what is observed in neutron scattering experiments. Data on
BaFe$_{1.7}$Ni$_{0.3}$As$_2$, which corresponds to nominally 24.6 electrons per unit
cell, shows that a spin gap remains open and the spectral weight has shifted
upward to 60 meV \cite{WZLT13}. In our calculations in the 16-orbital model, we find a spin
gap of about 30 meV at 0.4 electrons is pushed upwards to nearly 60 meV at
0.475 electrons. This agreement suggests that the low-energy dynamics may be
well represented in our model.

Moreover we note that, experimentally, the critical doping $x\approx 0.25$ at
which $T_c$ goes to zero in BaFe$_{2-x}$Ni$_x$As$_2$ corresponds to nominally
24.5 electrons per unit cell. Thus we find within the 16-orbital model that
the $s^\pm$ state is the leading pairing state over almost the full
superconducting dome in Ni-doped BaFe$_2$As$_2$.  We further note that a full
theory that takes into account the dynamics of the interaction would pick up
the spectral weight in the spin fluctuation spectrum at higher energies  and
thus likely extend the doping range over which the $s^\pm$ state is dominant
\cite{MiSM16,LMWJ16}.

In order to better understand how the arsenic $p$-orbitals give rise to this
behavior, we calculate the orbital contributions to the the Fermi surface
Bloch states,
\begin{align} 
	\label{eq:orbWeights} w_i^\ell = \int_{{\cal C}_i} \frac{d{\bf
	k}}{(2\pi)^2} \left| a^\ell_\mu({\bf k}) \right|^2 \,. 
\end{align}  
Here the integral is over the Fermi surface momenta of sheet $i$ and $\mu$ is
the band index of the sheet. 

\begin{figure} [t] 	
		\includegraphics[scale=0.42]{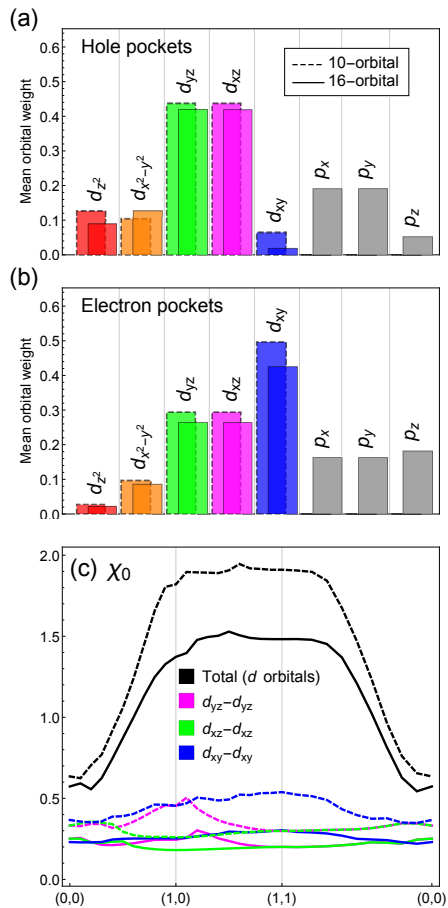}
		\caption{ (a),(b) Orbital weights of the Fermi surface Bloch states for 12.4 electrons per two
		iron atoms (overdoped; BaFe$_{1.8}$Ni$_{0.2}$As$_2$) for the 10-
		(dashed borders) and 16-orbital models (solid borders). (a): Orbital
		weights for the two $\Gamma$-centered hole pockets. (b): Orbital weights for the two electron pockets.  The
		reduction in $d_{xy}$ intensity on the electron pockets leads to a
		reduction of the spin susceptibility near the wavevector $(\pi,\pi)$
		connecting them. (c)  Bare (noninteracting) susceptibility $\chi_0$
		and intra-orbital $\chi_0^{\ell_1,\ell_1,\ell_1,\ell_1}$ for
		$\ell_1=d_{xz}$, $d_{yz}$, and $d_{xy}$, for the 10-orbital (dashed)
		and 16-orbital model (solid). } \label{fig4}
\end{figure}

In Fig.~4 we plot the orbital weights $w^\ell_i$ for both the 10-orbital
(dashed outline) and the 16-orbital (solid outline) model summed over the
different hole pockets (a) and electron pockets (b).
In the 16-orbital model, one observes that the arsenic $p$-orbitals hybridize
with all of the iron $d$-orbitals, reducing the orbital content of all five
$d$-orbitals on the Fermi surface states. The most significant change in terms
of total orbital weight is the reduction of the $d_{xy}$ content on the
electron pockets. Also shown in Fig.~4 are the largest intra-orbital
contributions ($\ell_1=\ell_2$) to the spin susceptibility
$\chi^{0}({\bf q},\omega=0)$ in Eq.~\eqref{eqn:chi0} for both the
10-orbital and 16-orbital models.  From this one sees that the $(\pi,0)$
peak in $\chi^{0}({\bf q},\omega=0)$ arises mainly from scattering
between the $d_{yz}$ orbitals on the hole and electron pockets, while the
$d_{xy}$ contribution is dominant in the near $(\pi,\pi)$ scattering. Relative
to the 10-orbital model, the $d_{xy}$ scattering is reduced more than the
$d_{yz}$ scattering, consistent with the larger reduction of the $xy$ orbital
weights on the electron pockets. Hence, it is the fact that the arsenic
$p$-orbitals have the strongest hybridization with the $d_{xy}$ orbitals on
the electron pockets that leads to the reduction of the near $(\pi,\pi)$
spin-fluctuation scattering and ultimately to the increased stability of the
$s^\pm$ pairing state with electron doping. 

\section{Conclusions}

To summarize, we have carried out RPA spin-fluctuation calculations of the
spin susceptibility and the leading pairing states in a 16-orbital
Hubbard-Hund tight-binding model of electron-doped BaFe$_2$As$_2$. In addition
to the 10 Fe-$d$ orbitals per unit cell, this model includes the six $p$-orbitals from the As atoms and their hybridization with the $d$-orbitals. We
have compared the results of these calculations with those of a Fe-$d$,
10-orbital only model that does not include the As-$p$ degrees of freedom. In
both models we find a leading $s^\pm$ pairing state and a subleading $d_
{x^2-y^2}$-wave state in the parent compound. Upon doping, the 10-orbital
model has the $d$-wave state become the leading state already at an electron
doping level of less than 0.2 electrons per unit cell. In contrast, in the
16-orbital model the $s^\pm$ state is much more stable and remains the leading
pairing state and dominant over the $d$-wave state up to doping levels of
0.475 electrons. At this doping level, the hole Fermi surface pockets at the
zone center are found to disappear. The increased stability of the $s^\pm$
over the $d$-wave state is found to arise from an increased ratio of the
strength of spin-fluctuation scattering near $q= (\pi,0)$ which connects hole-
and electron pockets and $q=(\pi,\pi)$ which connects the electron pockets.
This reduction of the near $(\pi,\pi)$ scattering is found to be unrelated to
a change in the Fermi surface shape, which is nearly identical between the 10-
and 16-orbital models, but rather can be traced to a decrease in the $d_{xy}$
orbital weight on the electron pockets due to their hybridization with the
As-$p$ degrees of freedom.

Orbital selectivity, i.e. the orbital dependent coherence of quasiparticles,
has been argued to play a major role in the Cooper pairing in iron-based
superconductors \cite{Yin11,Medici14,Kreisel17,Sprau17,LiLifeas}. In particular, Kreisel et
al.  \cite{Kreisel17} have found that the incorporation of quasiparticle weight factors can
modify the results of RPA spin-fluctuation calculations of the pairing state
via suppression of the pair scattering processes involving the less coherent
$d_{xy}$ states. Here we have shown that the orbital dependent hybridization
of Fermi surface Bloch states with the usually neglected $p$-orbital states
provides another, complementary ingredient for an improved itinerant pairing
theory.

\section*{Acknowledgements}
The RPA calculations in this work have been supported by NSF under Grant No. NSF-DMR-1308603 (D.W.T.). The analysis and interpretation of the results (T.A.M.) and the DFT and Wannier function calculations (T.B.) were supported by the U.S. Department of Energy, Office of Basic Energy Sciences, Materials Sciences and Engineering Division.

\end{document}